\documentclass[aps,prl,showpacs,twocolumn,superscriptaddress,floats,
floatfix]{ revtex4}
\usepackage{mathrsfs}
\usepackage{graphicx}
\usepackage{amsmath,amssymb,amsfonts}

\begin{document}

\title{Active elastic dimers: self-propulsion and current reversal on a
featureless track}

\author{K. Vijay Kumar}
\email{vijayk@physics.iisc.ernet.in}
\affiliation{Centre for Condensed Matter Theory, Department of Physics,
Indian
Institute of Science, Bangalore 560012, India.}

\author{Sriram Ramaswamy}
\email{sriram@physics.iisc.ernet.in}
\altaffiliation[Also at~]{CMTU, JNCASR, Bangalore 560064, India.}
\affiliation{Centre for Condensed Matter Theory, Department of Physics,
Indian Institute of Science, Bangalore 560012, India.}

\author{Madan Rao}
\email{madan@rri.res.in}
\altaffiliation[Also at~]{NCBS (TIFR), Bangalore 560065, India.}
\affiliation{Raman Research Institute, Bangalore 560080, India.}

\date{\today}

% ABSTRACT
\begin{abstract} 
We present a Brownian inchworm model of a
self-propelled elastic dimer in the absence of an external potential.
Nonequilibrium noise together with a stretch-dependent damping form the
propulsion mechanism. Our model connects three key nonequilibrium
features -- position-velocity correlations, a nonzero mean
internal force, and a drift velocity. Our analytical results, including 
striking current reversals, compare
very well with numerical simulations.  The model unifies the propulsion 
mechanisms of DNA helicases, polar rods on a vibrated surface, crawling 
keratocytes and Myosin VI. We suggest experimental realizations and tests
of the model.  
\end{abstract}


\maketitle

%%%%%%%%%%%%%%
% INTRODUCTION
Directed motion without an imposed external gradient in a homogeneous,
isotropic environment is seen not only in living systems \cite{bray}
but also in agitated granular matter \cite{yamada,kudrolli05}.  Can
these apparently diverse systems be understood in a unified manner?
We argue here that they can, and present a model which applies,
suitably interpreted, to the movement of helicases on DNA
\cite{schulten_biophys_06}, the directed motion of macroscopic polar
rods lying on a vertically vibrated surface \cite{yamada}, the
crawling of keratocytes which contain treadmilling actin
\cite{verkhovsky_kerato_currbiol} and the walking of processive motors
\cite{vale} such as Myosin VI on actin filaments
\cite{altman_sweeney_spudich_cell_2004}. While not losing sight of
the application to particular organisms or devices, our focus is on
the general principles governing propulsion by rectification of an
unbiased 
input active noise in a homogeneous medium. 

In all the systems mentioned above, macroscopic directed motion of the
center-of-mass (CM) arises via a coupling to internal coordinates, as a
result of two crucial features -- an asymmetrical environment for the
internal coordinates and external energy input. Unlike in traditional
``\mbox{Brownian ratchet models}'' \cite{julicher_reimann} of
directed motion, the asymmetry of interest in the above systems is
\emph{internal} to the motile objects, and does not lie 
in an external periodic potential. Our approach is distinct from that of
\cite{jayannavar} where the external
potential plays a central role, and also differs from the dynamical
systems approach of \cite{denisov}. The present model is 
similar in spirit to \cite{mogilner,norden} but simpler, 
and differs in several important details as seen below. We find 
an unexpected range of possible behaviors, especially 
in the dependence of the motion on the details of the nonequilibrium noise.

%%%%%%%%%%%%%%%%%%%%%%%%%%%%
% SUMMARIZE MODEL AND RESULTS
Our model self-propelled object is a dimer whose two
heads are coupled by a spring, in a homogeneous, dissipative, noisy
environment. The damping coefficients of the heads depend on the
relative coordinate or \emph{strain}.  The noise on the particles is
made of two parts -- a \emph{thermal} part whose strength is
determined by a fluctuation-dissipation relation with the
strain-dependent damping, and a nonequilibrium or \emph{active} part,
with strength independent of the damping, which represents
the external energy input. 

Our results are as follows: 
(i) The steady-state average of the CM velocity is in general nonzero and
exhibits counter-intuitive 
reversals of direction as a function of the strengths and characteristics 
of the drive and the dampings.
(ii) The steady state displays two other key nonequilibrium features: 
the mean internal force as well as the the equal-time correlation of the
relative 
coordinate to the CM velocity are both nonzero. 
(iii) Active noise \emph{alone} will not result in propulsion, even in
the presence of an asymmetric internal potential; the strain-dependent 
damping is an essential ingredient. 
(iv) The preceding results, obtained by perturbative analytical solution of
our 
model Langevin equations, with the coefficient of the stretch-dependent
damping as a small parameter, are confirmed in detail by direct numerical 
solution of the equations of motion of the particles.

%%%%%%%%%%%%%%%%
% DESCRIBE MODEL
The heads of the dimer are two point masses $m_i$, $i=1,2$,
with positions $x_i(t)$ at time $t$ and relative coordinate $x \equiv
x_1-x_2$ connected by a spring potential $U(x)$, with a minimum at
$x_m$, and acted upon by viscous damping and noise with a
nonequilibrium component.  Thus the Langevin equations of the
particles, in the It$\Hat{\textrm{o}}$ interpretation, are
\begin{equation} 
m_i \dot{v}_i + \alpha_i(x) v_i = 
	-\partial_{i} U + \sqrt{2 \alpha_i(x) k_B T} \, f_i + \sqrt{A_i} \,
\zeta_i 
\label{eq:x1x2_eom}	
\end{equation} 
where $v_i$ are the velocities of the two heads, the overdot
indicates a time-derivative, $\partial_i \equiv \partial / \partial
x_i$, $\alpha_i(x)$ are damping coefficients which depend on the
internal coordinate $x$, the terms containing the unit-strength,
independent Gaussian white noise sources $f_i(t)$ and $\zeta_i(t)$
encode thermal and nonequilibrium agitation respectively, $k_B$ and
$T$ are Boltzmann's constant and the thermodynamic temperature, and
$A_i$ is a measure of the external energy input. We consider
stretch-dependent dampings with $\alpha_i(x) > 0$ to ensure positive
dissipation.  For calculational convenience and concreteness we
consider smooth spring potentials, but the qualitative results of our
model hold for any confining $U$.

%%%%%%%%%%%%%%%%%%%%%%%%%%%%%%%%%%%%%%%%%%%%%%%%%%%%%%
% SYMMETRY CONSIDERATIONS, RELATION TO INTERNAL STRESS
Before solving (\ref{eq:x1x2_eom}), some general features are worth
noting. 
The noise-averaged internal velocity 
$\langle \dot{x} \rangle = 0$ in the steady state,  
as long as $U$ confines $x$ so that such a steady state 
exists. Now consider the special case where the $\alpha_i$ are independent
of $x$. 
By inspection of (\ref{eq:x1x2_eom}), we see then that the individual
velocities 
$\langle v_i \rangle = 0$ even for a non-centrosymmetric $U(x)$, 
despite the presence of the nonequilibrium noises $\zeta_i$, 
and even for the ``two-temperature'' \cite{schmittmann_zia}  
case $A_1 / \alpha_{1} \neq A_2 / \alpha_{2}$. Stretch-dependent damping 
is crucial to produce drift of the CM coordinate in this model.   

For a stiff enough spring, the dimer will explore small values of $x$ so
that 
$\alpha_i(x) \simeq \gamma_{0} + \gamma_{i} x$ where, for 
simplicity and with only trivial loss of generality, we have taken 
the $x$-independent part of the dampings on the two heads to be equal.
Averaging over the noise in 
(\ref{eq:x1x2_eom}), we find
\begin{equation}
\frac{ \langle v_i \rangle}{( \gamma_{1} + \gamma_{2})} 
	= \frac{\langle U'(x) \rangle}{\gamma_{0} (\gamma_{1} - \gamma_{2}) }
	= - \frac{\langle x v_i \rangle}{2 \gamma_{0}}
\label{eq:connecting_equation}
\end{equation}
where the prime denotes differentiation with respect to $x$.
This relation connects three key quantities --  
mean drift velocity, correlation of internal coordinate and drift
velocity, 
and mean internal force -- each of which can be nonzero only away from
thermal 
equilibrium. In particular, we see that a nonzero mean internal
force (a force dipole \cite{aditisr2002}) 
is linked to $x v_i$ correlations, and that either leads to 
drift if the damping is stretch-dependent ($\gamma_{i} \neq 0$). The above
relation also elucidates the manner in which an internal asymmetry in 
$x$ leads to macroscopic directed motion: the drift velocity and the
internal coordinate are correlated in the presence of a nonequilibrium
driving force.

%%%%%%%%%%%%%%%%%%%%%%%%%%%%%%%%%%%%%%%%%
% RELATION TO OTHER SELF-PROPELLED OBJECTS
Before analyzing our model in detail we examine four examples where
internal frictional asymmetry and nonequilibrium noise lead to directed
motion. 

(i) Structural studies \cite{velankar} together with the findings of a
recent
molecular simulation \cite{schulten_biophys_06} of the PcrA helicase
motor on single stranded DNA are of particular interest: its protein
domains 1A and 2A contract around ATP and catalyze its hydrolysis which 
then actively stretches them apart. Stretch-dependent damping as in our Eq.
(\ref{eq:x1x2_eom}) is encoded in the fact that in the ATP-bound (free)
state 1A has a higher (lower) barrier to motion than 2A. The periodic
potential in
\cite{schulten_biophys_06} is centrosymmetric, and serves only to provide
the
barriers that define the mobility. Holding the relative coordinate out of
equilibrium, in this case by maintaining a disequilibrium between ATP and
ADP $+$ P$_\textrm{i}$, results in motion in the direction of 2A. 

(ii) Polar granular rods on a vertically vibrated, horizontal plate were
studied in \cite{yamada}. The two ends of the rod have different friction,
so damping depends on the tilt, which is the internal coordinate of
interest. The nonequilibrium agitation being uncorrelated to this
frictional asymmetry, the CM of the rod translates. 

(iii) In the crawling of cells or cell-fragments driven by
``treadmilling actin'' \cite{verkhovsky_kerato_currbiol}, 
the relevant internal coordinate is the instantaneous degree of
polymerization, averaged over all filaments. This quantity is maintained in
a nonequilibrium steady state by the balance between ATP-aided
polymerization at the leading edge, and passive depolymerization at the
trailing edge.
This leads to more focal adhesions at the front of the cell. The stretched
cell thus detaches primarily at the rear, resulting in net translatory
motion. 

(iv) In the motion of Myosin VI \cite{altman_sweeney_spudich_cell_2004} the
detachment of the forward head is inhibited by the extension of the linker
connecting it to the rear head. 
The differential binding of the forward head depending on the extension of
the linker mimics our 
stretch-dependent damping, and ATP hydrolysis provides the energy source. 

Our model thus provides a unifying understanding of four quite distinct
self-propelled systems.

%%%%%%%%%%%%%%%%%%%%%%%%%
% QUALITATIVE DESCRIPTION
To understand qualitatively how the dimer walks, 
consider the case where the nonequilibrium noise and the stretch-dependent
damping 
(with a simple form interpolating smoothly between $\Gamma$ for $x > 0$
and 
$\Gamma' < \Gamma$ for $x < 0$) act only on one head of the dimer, say
particle 1. 
Let particle 2 have a fixed damping coefficient lying between $\Gamma$ and
$\Gamma'$. Suppose the active noise consists of discrete dimer-stretching 
events separated by intervals 
whose mean $\tau$ is much larger than the relaxation time $\tau_d$ of the
dimer. Then, if noise compresses (stretches) the dimer, particle 1 retracts
faster (slower) than
particle 2, leading to translation of the CM in the direction of particle
1. 
Suppose instead the active noise consists of a succession of small 
impulses, at intervals $\tau \ll \tau_d$ (effectively white noise 
as in this paper), then the dimer is kicked many times before it can relax.
Head 1, 
whose damping increases with stretch, will accumulate a smaller
displacement than head 2, 
so that the net displacement will be in the direction of head 2. This
latter 
behavior is in fact what our calculations and numerical studies find.
A purely equilibrium thermal white noise, with variance equal to $2k_BT$
times 
the stretch-dependent damping, will of course fail to produce any net
motion, 
because increases in the noise amplitude compensate precisely for the
enhanced damping.

%%%%%%%%%%%%%%%%%%%%%%%%%%%%%%%%%%
% EXPLAIN PERTURBATIVE CALCULATION
Analytical expressions for the inchworm speed, the average internal force
and related statistical descriptors of the motion can be obtained in a
perturbative treatment, 
expanding the damping coefficients to leading order in $x$. The value of
this approach is that 
it elucidates the separate and essential roles of the $x$-dependent damping
and the nonequilibrium noises, 
and shows the connection of the mean drift speed to the mean internal force
and 
correlations of $x$ with the CM velocity $V$. We choose $m_i=m$ and take a
harmonic internal potential 
$U(x) = \frac{1}{2} a x^2$. This simplifies considerably the perturbation
theory calculations 
that follow, without losing any of the essential physics. 
We also assume that the active noise is absent for $t<0$ and that at
$t=0$, 
the variables $x$, $\dot{x}$ and $V$ are at equilibrium at temperature $T$.
A formal solution for $V(t)$ and $x(t)$ can be written, using the
propagators from the linearized
version of the equations~(\ref{eq:x1x2_eom}), treating all the
nonlinearities as source terms, 
from which we approximate the solutions to successive orders in
perturbation
theory. Since we are interested in \emph{velocities} and their
correlations 
with position coordinates, we find it convenient to retain inertia,
particularly 
in the numerical calculation, so as to avoid ambiguities in solving our 
stochastic differential equations~\cite{LauLubensky}. Our numerical studies are, however,
entirely in the overdamped regime.

%%%%%%%%%%%%%%%%%%%%%%%%%%%%%%
% EXPLAIN NUMERICAL SIMULATION
Numerical simulations were performed using an Euler-Maruyama 
scheme~\cite{kloeden_platen_schurz}. For simplicity, we take $x$-dependent
damping on 
only one of the particles, say the particle 1, while the other particle has
a constant 
value of the damping coefficient ($\gamma_2= 0$). Specifically, we choose 
$\alpha_1(x) = \frac{1}{2}\big[ (\Gamma+\Gamma') + (\Gamma-\Gamma')
\tanh(x/w)\big]$, where $w$ is the 
width over which the strain-dependent damping changes over between the two
extreme values of 
$\Gamma$ and $\Gamma'$. In this particular form, we have
$\gamma_{0}=(\Gamma+\Gamma')/2$ and 
$\gamma_{1}=(\Gamma-\Gamma')/2w$. We scale masses, lengths and times by 
$m$, $x_m$ and $m/\gamma_{0}$ respectively and use a constant dimensionless
time step 
of $\Delta t = 10^{-2}$. Averages are reported over $n=10^6$ realizations
of the noise.
We choose plausible values of $a=0.05$, $k_B T=0.01$ based on those quoted
in 
\cite{schulten_biophys_06} for the dimeric PcrA helicase and take
$(\Gamma+\Gamma') = 2$ 
in dimensionless units. To vary $\gamma_{1}$, we fix a value of $w$ and
change $(\Gamma - \Gamma')$.

%%%%%%%%%%
% RESULTS
The linearized version of the equations (\ref{eq:x1x2_eom}), retaining  
the nonequilibrium noise, can be solved exactly. To this order, the noise
averaged steady state 
values of the CM velocity and the internal force vanish. 
However, the equal time correlators have a distinct contribution from the 
nonequilibrium noise. In particular, the steady state value of the equal
time
correlator between the internal coordinate and the CM velocity is
\begin{equation}
\label{eq:xV0}
\langle x V \rangle_0 = \frac{(A_1 - A_2)}{ 4 ( \gamma_{0}^2 + a m )}.
\end{equation}
where the subscript $0$ indicates the order in perturbation theory. 
Notice that $\langle x V \rangle_0$ is proportional to 
$\Delta T_{neq} \equiv (A_1 - A_2) / \gamma_{0}$, which has a precise 
interpretation as the difference in the effective temperatures of the
nonequilibrium
noise terms of the two heads constituting the dimer. At thermal
equilibrium, by contrast, 
all equal-time correlations between $x$ and $V$ must vanish. 
The nonzero correlation (\ref{eq:xV0}) between $x$ and $V$ leads to 
nonzero steady-state averages for $V$ and $x$ at first order in
perturbation theory 
through (\ref{eq:connecting_equation}). Note from
(\ref{eq:connecting_equation}) and (\ref{eq:xV0}) 
that a nonzero mean drift velocity and mean internal force require
$\gamma_{i} \neq 0$, 
i.e., stretch-dependent friction, in addition to the two-temperature
scenario just described.

\begin{figure}
  \begin{center}
		\includegraphics[width=0.85\columnwidth]{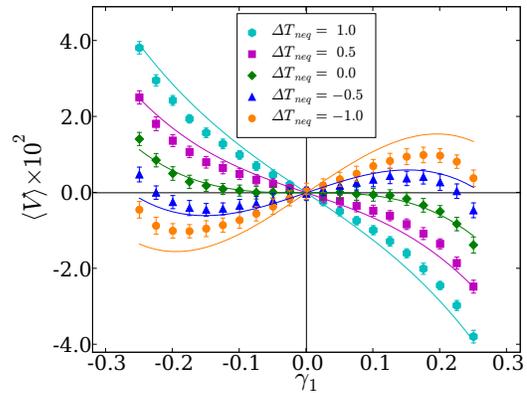}
		\caption{\label{fig:current_reversal}{\small (Color online)} 
	CM velocity as a function of the stretch dependent damping coefficient 
	$\gamma_{1}$ for various $\Delta T_{neq}$ with $\gamma_2=0$ and $w=4.0$. 
	$\Delta T_{neq}$ was changed by fixing 
	the value of $A_1=1.0$  and varying $A_2$. The solid lines are the
corresponding perturbation 
	theory calculations from equation (\ref{eq:xV2}).}
  \end{center}
\end{figure}

From equation (\ref{eq:connecting_equation}), we notice that since $x \to
-x$ and $v_i \to -v_i$ 
is a symmetry of the equation, $\langle xv_i \rangle$ must be even in the
$\gamma_i$. 
A calculation of the equal time correlator $\langle x V \rangle$ to next
order in 
perturbation theory yields 
\begin{eqnarray}
\langle x V \rangle_2 &=& C_0 \, (\Delta T_{neq})
		+  C_1 \, (\Delta T_{neq})^2 \, (\gamma_{1}+\gamma_{2})^2
		\nonumber \\
		&+&  C_2 \, (\Delta T_{neq}) \, A \, (\gamma_{1}-\gamma_{2})^2
		\nonumber \\
		&+& C_3 \, (\Delta T_{neq})^2 (\gamma_{1}^2-\gamma_{2}^2)
		+  C_4 A^2 (\gamma_{1}^2 - \gamma_{2}^2)
\label{eq:xV2}
\end{eqnarray}
where $A = A_1 + A_2$ and the $C_i$ are coefficients depending on $a$,
$\gamma_0$ and $m$. Notice from 
the above equation that if both $\gamma_1 = \gamma_2$ and $\Delta T_{neq} =
0$, $\langle x V \rangle$ 
vanishes and hence there can be no motion of the CM coordinate at this
order. 
The slope of the $V$ vs. $\gamma_1$ curve at the origin is 
proportional to the value of $\Delta T_{neq}$. Tuning the value of $\Delta
T_{neq}$ changes the slope and 
hence the small $\gamma_1$ dependence of $V$. However at large $\gamma_1$,
this effect becomes small 
and the $V$ goes as a power of $\gamma_1$. Thus there must be a current 
reversal at some $\gamma_1$. Equation (\ref{eq:xV2}) indeed predicts
current reversals 
as a function of the $\gamma_i$ depending on the values of the $A_i$. 
Numerical simulation results confirm this (Fig. \ref{fig:current_reversal})
and the current 
reversals agree reasonably with perturbation theory calculations.

%%%%%%%%
% FIGURE

With the increase in $A_i$, the CM velocity does not saturate in our simple
model. 
This is because the harmonic spring can stretch indefinitely and thus the
dimer can be 
in an infinite number of states defined by the relative coordinate $x$. A
real molecular motor,
on the other hand, cannot consume an indefinite number of ATP units and
thus its velocity saturates 
with the increase in ATP concentration. Modifying $U(x)$ so that 
the effective range of $x$ is limited should give rise to a saturation of
the CM velocity 
with increase in the input energy. 

In the presence of an external load we find a linear relation, to lowest
order in perturbation 
theory, between the steady state CM velocity and the applied force. The
reason for a linear relation is 
that the applied external force does not alter the mechanism of energy
uptake in our simple model.

We also find a generalized efficiency \cite{derenyi_99} $\eta \approx 2 m
\gamma_{0} \langle V \rangle^2/A$ 
to leading order in perturbation theory, in the absence of an external
load. With $\gamma_{1}=0.1$, $\gamma_2=0$ and $A_1 = 1.0$, $A_2 = 0.0$, we
find $\eta \approx 0.2\%$.

The results above are for the case of a harmonic spring potential
and a non-centrosymmetric $\alpha_i(x)$. A non-centrosymmetric
$U(x)$ and a centrosymmetric $\alpha_i(x)$ also yields directed motion.
Crucially, even in this case, the damping must be $x$-dependent for a
nonzero drift velocity. A non-centrosymetric $U(x)$ alone
does not lead to nonzero $\langle V \rangle$. We have confirmed this with
an
explicit numerical simulation.

%%%%%%%%%%%%%
% DISCUSSION
A few words comparing our model to that of \cite{mogilner}. 
Although the idea of stretch-dependent damping is present in these papers,
our model 
is much simpler in the way the separation of equilibrium and nonequilibrium
forces are 
presented. In \cite{mogilner}, stretch dependent damping is \emph{induced}
by activity 
whereas in our model it is present even in the absence of the
nonequilibrium driving noise. 
It does not lead to directed motion because of the the strength of the
equilibrium noise exactly 
balances the dissipation according to the fluctuation-dissipation theorem.
Directed motion is 
induced by the acive noise whose strength is independent of the
$\alpha_i(x)$. 
Also in our model, equation (\ref{eq:connecting_equation}) explicitly
clarifies the manner in 
which asymmetry in an internal degree of freedom is coupled to a
macroscopic coordinate in the 
presence of nonequilibrium noise, and leads to directed motion. That this
is also proportional to a 
nonzero average internal force highlights the nonequilibrium nature of the
phenomenon. 

A likely realization of this model is \cite{ramin} in the form of a
colloidal bead with a polymer tail. 
The damping on the CM of the polymer will depend significantly on its
stretch, while that on the bead will not. Subjecting this composite colloid
to nonequilibrium noise (chemical reactions, catalysis at its surface
\cite{kapral}, fluctuating laser interference patterns) 
should cause it to drift in the direction of its instantaneous
orientation. 
One-dimensional versions could be constructed using optical tweezers in a 
line trap geometry. This system could serve to test our theory.  

Tests should focus on (\ref{eq:connecting_equation}) as well as the
phenomenon of current reversal (Fig. \ref{fig:current_reversal}). In our
model the dampings and their $x$-dependences are properties intrinsic to
the dimer in the absence of active noise. Changing the strength of the
active noise should change the numerators in
(\ref{eq:connecting_equation}), leaving the denominators unchanged, thus
allowing a test of the two equalities in (\ref{eq:connecting_equation}).
Current reversals (Fig. \ref{fig:current_reversal}) are best probed by
altering the active noise 
levels on each head of the dimer. 

Several natural generalizations of our model suggest themselves, 
and will be investigated in the near future. 
These include Poisson or other active noises,
dimers moving in more than one dimension, 
coupled arrays or an elastic continuum of active particles \cite{joannyetal}, and coupling to hydrodynamic flow.

%%%%%%%%%%%%%%%%%
% ACKNOWLEDGMENTS
\begin{acknowledgments}
We thank P.~Perlekar, A.~M.~Jayannavar, D.~Lacoste and K.~Schulten for fruitful
discussions. The Centre for Condensed Matter Theory is supported by the 
DST, India. SR and MR acknowledge support from IFCPAR grant 3504-2. 
SR acknowledges a J.~C.~Bose Fellowship of the DST, India. This research was supported in 
part by the National Science Foundation under Grant No. PHY05-51164, and by the ICMR,
UCSB.

\end{acknowledgments}

%%%%%%%%%%%%%%%
% BIBLIOGRAPHY

\end{document}